
\input phyzzx

\REF\PKT{P.K. Townsend, {\it The 11-dimensional supermembrane
revisited}, preprint hepth-9501068, Phys. Lett. {\bf B}, {\sl in
press}.}
\REF\BST{E. Bergshoeff, E. Sezgin and P.K. Townsend, Phys. Lett. {\bf 189B}
(1987)
75; Ann. Phys. (N.Y.) {\bf 185} (1988) 330.}
\REF\HS{G. Horowitz and A. Strominger, Nucl. Phys. {\bf B360} (1991) 197.}
\REF\W{E. Witten, {\it String theory dynamics in various dimensions}, preprint
hepth-9503124.}
\REF\DHIS{M.J. Duff, P.S. Howe, T. Inami and K.S. Stelle,
Phys. Lett. {\bf 191B} (1987) 70.}
\REF\DS{M.J. Duff and K.S. Stelle, Phys. Lett. {\bf B253} (1991) 113.}
\REF\DGT{M.J. Duff, G.W. Gibbons and P.K. Townsend, Phys. Lett. {\bf 332 B}
(1994) 321.}
\REF\GHT{G.W. Gibbons, G.T. Horowitz and P.K. Townsend, Class. Quantum Grav.
{\bf
12} (1995) 297.}
\REF\DHGR{A. Dabholkar, G.W. Gibbons, J.A. Harvey and F. Ruiz-Ruiz, Nucl. Phys.
{\bf B340} (1990) 33.}
\REF\G{R. G\"uven, Phys. Lett. {\bf 276B} (1992) 49.}
\REF\HT{C.M. Hull and P.K. Townsend, Nucl. Phys. {\bf B438} (1995) 109.}
\REF\BPS{I. Bars, C.N. Pope and E. Sezgin, Phys. Lett. {\bf 198B} (1987) 455.}
\REF\PS{C.N. Pope and K.S. Stelle, Class. Quantum Grav. {\bf 5} (1988) L161.}
\REF\dWN{B. de Wit and H. Nicolai, {\it Supermembranes - a fond farewell?} in
{\sl Supermembranes and Physics in 2+1 Dimensions}, eds. M.J. Duff, C.N. Pope
and
E. Sezgin (World Scientific, 1990).}
\REF\HTb{C.M. Hull and P.K. Townsend, in preparation.}
\REF\DL{M.J. Duff and J.X. Lu, Nucl. Phys. {\bf B354} (1991) 141.}
\REF\CHS{C. Callan, J.A. Harvey and A. Strominger, Nucl. Phys. {\bf B359}
(1991) 611.}
\REF\DH{A. Dabholkar and J.A. Harvey, Phys. Rev. Lett. {\bf 63} (1989) 719.}
\REF\vNT{P.K. Townsend and P. van Nieuwenhuizen, Phys. Lett. {\bf 125B} (1983)
41; L. Mezincescu, P.K. Townsend and P. van Nieuwenhuizen, Phys. Lett. {\bf
143B}
(1984) 384.}
\REF\BKS{E. Bergshoeff, I.G. Koh and E. Sezgin, Phys. Rev. {\bf D32} (1985)
1353.}
\REF\K{R.R. Khuri, Phys. Lett. {\bf  B259} (1991) 261; Nucl. Phys. {\bf B387}
(1992) 315.}
\REF\GHL{J. Gauntlett, J. Harvey and J. Liu, Nucl. Phys. {\bf B409}
(1993) 363.}
\REF\S{A. Sen, Int. J. Mod. Phys. {\bf A8} (1993) 5079.}
\REF\NT{H. Nicolai and P.K. Townsend, Phys. Lett. {\bf 98B} (1981) 257.}
\REF\DNP{M.J. Duff, B.E.W. Nilsson and C.N. Pope, Phys. Lett {\bf 129B}
(1983) 39.}
\REF\D{M.J. Duff, {\it Strong/Weak Coupling Duality From the Dual String},
preprint hep-th/9501030.}
\REF\Sb{A. Sen, {\it String-String Duality Conjecture in Six Dimensions and
Charged Solitonic Strings}, preprint hep-th/9504027.}
\REF\HStr{ J.A. Harvey and A. Strominger, {\it The Heterotic String is a
Soliton}, preprint hep-th/9504047.}
\REF\GT{G.W. Gibbons and P.K. Townsend, Phys. Rev. Lett. {\bf 71} (1993) 3754.}
\REF\HST{P.S. Howe, G. Sierra and P.K. Townsend, Nucl. Phys. {\bf B221} (1983)
331.}
\REF\PKTb{P.K. Townsend, Phys. Lett. {\bf 309B} (1993) 33.}
\REF\PKTc{P.K. Townsend, Phys. Lett. {\bf 139B} (1984) 283.}
\REF\AETW{A. Ach\'ucarro, J. Evans, P.K. Townsend and D. Wiltshire,
Phys. Lett. {\bf 198B} (1987) 441.}
\REF\St{A. Strominger, Nucl. Phys. {\bf B343} (1990) 167.}
\REF\CHSb{C. Callan, J.A. Harvey and A. Strominger, Nucl. Phys. {\bf B367}
(1991)
60.}
\REF\STo{G. Sierra and P.K. Townsend, Nucl. Phys. {\bf B223} (1984) 289.}
\REF\GR{S.J. Gates, Jr. and V.G.J. Rodgers, {\it A Truly Crazy Idea about Type
IIB Supergravity and Heterotic Sigma-Models}, preprint hep-th/9503237.}
\REF\DLb{M.J. Duff and J.X. Lu, Phys. Lett. {\bf 273B} (1991) 409.}
\REF\ABEQUIV {J. Dai, R.G. Leigh and J. Polchinski, Mod. Phys. Lett. {\bf
A4} (1989) 2073; M. Dine, P. Huet and N. Seiberg, Nucl. Phys. {\bf B322}
(1989) 301.}

\def\square{\kern1pt\vbox{\hrule height 0.5pt\hbox
{\vrule width 0.5pt\hskip 2pt
\vbox{\vskip 4pt}\hskip 2pt\vrule width 0.5pt}\hrule height
0.5pt}\kern1pt}


\Pubnum{ \vbox{ \hbox{R/95/12} \hbox{hep-th/9504095}} }
\pubtype{}
\date{April, revised May, 1995}

\titlepage

\title{String-Membrane Duality in Seven Dimensions}

\author{P. K. Townsend}
\address{DAMTP, University of Cambridge,
\break
Silver Street, Cambridge CB3 9EW,  U.K.}

\abstract {The conjectured equivalence of the heterotic string to a $K_3$
compactified type IIA superstring is combined with the conjectured
equivalence of the latter to a compactified 11-dimensional supermembrane to
derive a string membrane duality in seven dimensions; the membrane is a soliton
of the string theory and vice versa. A prediction of this duality is that the
heterotic string is a $K_3$ compactification of the solitonic 11-dimensional
fivebrane. It is verified that the worldsheet action of the D=10 heterotic
string
is indeed obtainable by $K_3$ compactification of the worldvolume action of the
11-dimensional fivebrane, and it is suggested how the worldvolume action of
the D=11 supermebrane may be similarly obtained by $T^3$ compactification of
the
worldvolume action of a D=10 heterotic fivebrane. Generalizations to $D=8$
string-threebrane and membrane-membrane duality are also discussed.}

\endpage

\pagenumber=2


\chapter{Introduction }

According to a recent proposal [\PKT], the ten-dimensional (D=10) type IIA
superstring theory is an $S^1$ compactification of the, still conjectural,
11-dimensional (D=11) supermembrane theory [\BST]. The impetus for this
proposal
came from the realization that, in the quantum theory, the extreme electric
black
holes of the D=10 N=2A supergravity theory [\HS] would be indistinguishable
from
a tower of Kaluza-Klein (KK) states. This observation has since been made
independently by Witten [\W] who has also pointed out that since the $S^1$
radius
goes to infinity with the string coupling constant, the strong coupling limit
of
the type IIA superstring theory can be viewed as a decompactification limit in
which 11-dimensional Lorentz invariance is restored. These observations are a
clear indication that the type IIA superstring theory has an 11-dimensional
origin, but they do not, by themselves, provide evidence for an 11-dimensional
supermembrane. Indeed, the only low mass excitations of the type IIA
superstring
in the strong coupling limit are the massless excitations of 11-dimensional
supergravity [\W]. The reason for the identification in [\PKT] of the
11-dimensional theory as a supermembrane theory, rather than merely D=11
supergravity, is to be found partly in earlier work identifying the type IIA
worldsheet action as a double dimensional reduction of the supermembrane
worldvolume action [\DHIS], also partly in work [\DS,\DGT,\GHT] relating
the `solitonic' string [\DHGR] and other p-brane solutions [\HS] of D=10 N=2A
supergravity to the membrane [\DS] and fivebrane [\G] solutions of
11-dimensional
supergravity, and in [\HT] where it was suggested, on the basis of U-duality in
type II string theory, that the `solitonic' membrane of D=11 supergravity be
identified with a fundamental supermembrane.

This proposal is consistent
with the observations of [\W] concerning the low mass excitations in the strong
coupling limit of the type IIA superstring if (i) the zero mass excitations of
the
supermembrane are those of D=11 supergravity, as advocated in [\PKT] partly on
the basis of earlier work [\BPS,\PS] on supermembranes\foot{But see [\dWN] for
a
review of the arguments against.}, and (ii) there are no other low mass
excitations, which would not be surprising in view of the absence in 11
dimensions of a dilaton field, and hence of any weak coupling limit.

According to another proposal [\HT], the $K_3$-compactified type IIA
superstring
is equivalent, non-perturbatively, to a toroidally compactified heterotic
string. The basis for this proposal was that (i) the low energy effective
supergravity theories are the same and (ii) the spectrum of massive
`Bogomolnyi'
states may also be the same once one includes `wrapping' modes of solitonic
p-brane solutions of the ten-dimensional theories; these are analogous to
string
winding states, but are non-perturbative. Thus, while the perturbative spectrum
of massive states is very different in the two theories, this difference could
disappear in the full theory. Whether this actually happens\foot{And evidence
that it does is accumulating [\W,\HTb].} depends on a number of factors, one of
which is whether the D=10 p-brane solutions in question are truly `solitonic'
and
have therefore to be included when determining the spectrum of a compactified
string theory. If `solitonic' is taken to mean that the `string sigma-model'
spacetime metric is geodesically complete and that all fields are everywhere
non-singular then, of the type IIA p-brane solutions, only the neutral
fivebrane
[\DL,\CHS] is truly `solitonic'. For the heterotic string theory this suffices
since the only other p-brane solution is the `solitonic' string and while this
is
not actually solitonic by the above criterion\foot{This is equally true in
terms
of the `fivebrane sigma-model' metric because the dilaton blows up at finite
affine parameter along timelike geodesics.}, this fact is unimportant if this
solution is identified with the fundamental string, as suggested in [\DH,\HT].
For the type IIA superstring, a similar justification of the necessity of
including p-brane solitons is possible only if they are re-interpreted as
solutions of D=11 supergravity and then only if the membrane solution is
identified with a fundamental supermembrane because, unlike the D=11 fivebrane,
the D=11 membrane solution has a (timelike) singularity behind its horizon;
this
fact was one of the principal motivations of [\PKT]. Thus the conjectured
equivalence of the heterotic string with the $K_3$ compactification of the type
IIA superstring is not independent of the identification of the latter as a
compactified supermembrane. It is therefore natural to consider these two
conjectures in conjunction with each other. This has already been done in [\W]
to
the extent that the strong coupling dynamics of the D=7 heterotic string was
related to a $K_3$ compactification of 11-dimensional supergravity. The purpose
of this paper is to consider the implications of supposing the strong coupling
limit of the D=7 heterotic string to be a $K_3$ compactified supermembrane. We
shall see that it implies a string/membrane duality in D=7.
One of the implications of this new duality is that the strongly coupled
$T^3$-compactified fundamental heterotic string has a solitonic interpretation
as
a $K_3$ compactified 11-dimensional fivebrane! As confirmation of this
prediction,
we shall verify that the physical worldsheet field content of the heterotic
string
is indeed found from the physical field content of the D=11 super-fivebrane by
compactification on $K_3$.

Conversely, according to D=7 string-membrane duality, the $K_3$-compactified
fundamental supermembrane has a solitonic interpretation as a
$T^3$-compactified
10-dimensional fivebrane. This prediction is not satisfied by the relevant
fivebrane solutions of D=10 supergravity/YM theory, but this field theory is
also not anomay free. It will be argued that the inclusion of the Lorentz
Chern-Simons terms needed for anomaly cancellation can lead to the desired
result. Finally, we discuss similar issues for the type D=10 IIB superstring in
the context of a possible generalization of D=7 string/membrane duality to D=8
string/threebrane duality.


\chapter{String/membrane duality}

Let us start with the $T^3$-compactified heterotic string. The effective D=7
field theory, at generic points in the moduli space of this compactification,
is
D=7 supergravity coupled to 19 vector supermultiplets. The bosonic field
content
of the supergavity multiplet is [\vNT]
$$
(g_{\mu\nu}, B_{\mu\nu}, V^a_\mu, \sigma)
\eqn\one
$$
where $V^a_\mu$, $(a=1,2,3)$, is an SO(3) triplet of vector potentials (which
remain abelian in the action of interest to us), and the scalar field $\sigma$
can be identified as the dilaton of string theory. The bosonic field content of
the D=7 vector multiplet is
$$
(V_\mu,\phi^a)
\eqn\two
$$
where $\phi^a$ is an SO(3) triplet of scalar fields. The coupling of an
arbitrary number, $k$, of vector multiplets to D=7 supergravity has been
constructed by Bergshoeff {\it et al} [\BKS]. Since the case of interest
here is $k=19$, there are a total of 22 vector fields $V_\mu^I$
$(I=1,\dots,22)$; let us denote by $F^I_{\mu\nu}$ their field strength
tensors. There are also a total of 58 scalar fields $(\sigma,\phi^i)$
$(i=1\dots, 57)$. The 57 scalars $\phi^i$ parametrize the sigma-model target
space
$$
{\cal M} = SO(3,19)/[SO(3)\times SO(19)]\ .
\eqn\three
$$
Let us denote by $m_{ij}$ the invariant metric on this coset space. After some
field redefinitions the bosonic action of [\BKS] can be rewritten as
$$
S=\int\!\! d^7x \; \sqrt{-g}\; e^{-2\sigma}\Big[ R +4(\partial\sigma)^2
-{1\over3}H^2 -a_{IJ}(\phi) F^IF^J
-2m_{ij}(\phi)\partial\phi^i\partial\phi^j\Big]
\eqn\four
$$
where $a$ is a positive-definite matrix function of the scalars $\phi$ which
can
be found in [\BKS], and the field strength tensor $H$ of the two-form potential
$B$ is given by
$$
H_{\mu\nu\rho} = 3\partial_{[\mu}B_{\nu\rho]} - A_{[\mu}^IF_{\nu\rho]}^J\;
\eta_{IJ}\ ,
\eqn\five
$$
where $\eta_{IJ}$ is the invariant $SO(3,19)$ tensor with diagonal entries
$(-1,-1,-1,1,\dots,1)$. It follows that the three-form $H$ satisfies the
modified
Bianchi identity
$$
dH = -{1\over3} F^I\wedge F^J\eta_{IJ}\ .
\eqn\six
$$

The action \four\ has been written in a form appropriate to its interpretation
as the effective action of the D=7 heterotic string. If we think of this
action as obtained by compactification from D=10 on $T^3$ then the two-form
potential $B$ obviously couples to the fundamental string. It also has a
magnetic `neutral' membrane source obtained by wrapping the D=10 neutral
fivebrane around the 3-torus. It is similarly straightforward to determine the
D=10 origin of the electric and magnetic sources of the 22 gauge potentials
$B_\mu^I$. Three of these are Kaluza-Klein vectors which couple to KK modes and
their (magnetic) threebrane duals (obtained by an obvious extension of the
standard KK monopole construction, as explained for the D=10 sixbrane in
[\PKT]).
Three more vector potentials arise from the D=10 two-form potential, and couple
to string winding modes and their threebrane duals (the obvious extension of
abelian H-monopoles [\K,\GHL]). The remaining 16 `heterotic' vectors couple to
the
charged modes of the additional 16 left-moving wordsheet scalars of
the heterotic string, and their magnetic duals, which are presumably \big(by
the
available D=4 evidence [\S]\big) threebrane generalizations of BPS monopoles.

The action \four\ has a dual version in which the two-form potential $B$ is
replaced by a three-form $A$ [\vNT]. Since the three-form field strength $H$
has
a modified Bianchi identity, the dual version will include a coupling of $A$ to
a topological current, as first discussed in [\NT]. Following the steps
presented
there, one finds that the dual action can be written as
$$
\eqalign{
\tilde S =\int\! d^7x\, \Big\{ &\sqrt{-g}\, e^{{4\over3}\sigma}\Big[ R
-{8\over3} (\partial\sigma)^2 -2m_{ij}\partial\phi^i\partial\phi^j -{1\over12}
F^2] \cr
&- \sqrt{-g}\, a_{IJ}F^JF^J + {1\over 18}\eta_{IJ}\;
\varepsilon^{\alpha\beta\gamma\mu\nu\rho\sigma}
A_{\alpha\beta\gamma}F^I_{\mu\nu}F^J_{\rho\sigma} \Big\}\ . }
\eqn\nine
$$
This is the effective D=7 action obtained by $K_3$ compactification of
11-dimensional supergravity [\DNP]. The last term can be seen to be a
direct consequence of the $\varepsilon AFF$ term in 11-dimensions and the fact
that $\eta_{IJ}$ is the intersection matrix of the 22 linearly independent
homology two-cycles of $K_3$. What we wish to consider here is the
11-dimensional
origin of the various massive modes or extended objects to which the seven
dimensional  gauge potentials couple. Let us concentrate first on the 22 vector
potentials. None of them are KK vectors since $K_3$ has no isometries. Instead,
they all derive from the ansatz
$$
A= B^I(x)\wedge \omega_I\ ,
\eqn\ten
$$
where the $\omega^I$ span the 22-dimensional space of closed but not exact
two-forms on $K_3$. If we suppose that the 11 dimensional supergravity is
merely
the effective action for the massless modes of a supermembrane then we must
also
include the membrane `wrapping' modes around the 22 two-cycles of
$K_3$. These are the electrically charged modes that couple to the 22 vector
potentials. Their magnetic duals are the 22 threebranes obtained by wrapping
the
D=11 fivebrane around the same 22 two-cycles. We now turn to the D=7 three-form
potential $A$. This is directly descended from the D=11 three-form, so
if we suppose that the latter couples to a fundamental 11-dimensional
supermembrane then it follows that the D=7 three-form also couples to a
fundamental supermembrane. Its magnetic source is the {\it solitonic} string
obtained by compactification of the D=11 fivebrane on $K_3$.

If we dualize the action \nine\ to recast it in the `heterotic' form of
\four\ then we have a two-form with a solitonic string source and a fundamental
magnetic membrane source, i.e. exactly the opposite of what we had before.
We therefore see that the dual interpretation of the D=7 Maxwell/Einstein
supergravity action {\it either} as a toroidally compactified D=10 heterotic
string {\it or} as a $K_3$ compactified D=11 supermembrane leads to a
string-membrane duality in D=7. The membrane appears as a soliton in the string
theory and the string appears as a soliton in the membrane theory. It seems
likely
that the string-membrane duality discussed here implies the string-string
duality
of [\D], for which further evidence has recently been found [\W,\Sb,\HStr].


\chapter{The heterotic string as a compactified D=11 fivebrane}

Perhaps the most interesting aspect of D=7 string/membrane duality is that,
from
the membrane perspective, the heterotic string theory is really a $K_3$
compactification of the D=11 super-fivebrane. It follows that the worldsheet
action of the heterotic string should be similarly related to the worldvolume
action of the D=11 fivebrane\foot{While this paper was nearing completion the
author became aware of the recent preprint of Harvey and Strominger [\HStr] in
which the same idea is developed.}. Although the latter is unknown, it {\it is}
known [\GT] that after (partial) gauge fixing it must be a supersymmetric field
theory based on the six-dimensional self-dual N=4 antisymmetric tensor
multiplet
[\HST], with field content
$$
(S^{[ij]}\, ;\, \lambda_+^i\, ;\, A^+\, )\ .
\eqn\eleven
$$
Here, $i=(1,2,3,4)$ is an index of $USp(4)\cong Spin(5)$, $\lambda_+^i$ is a
chiral `symplectic-Majorana' spinor of the six-dimensional Lorentz group and
$A^+$
is a two-form gauge potential with self-dual three-form field strength. The
scalars $S$ form a 5-plet of $Spin(5)$ and can be interpreted as the
Nambu-Goldstone fields due to the breaking of translational invariance in the
five transverse directions by the six-dimensional worldsheet of the D=11
fivebrane. In fact, the entire supermultiplet, including the antisymmetric
tensor
gauge potential, can be given a Nambu-Goldstone interpretation [\PKTb]. If we
fix
the antisymmetric tensor gauge invariance by the choice of lightcone gauge, so
reducing the supermultiplet \eleven\ to its purely physical field content, we
also
break the six-dimensional Lorentz group to $SO(1,1)\times [SU(2)\times
SU(2)]$. In the compactification of the fivebrane on some four-dimensional
space
$B$, the $SO(1,1)$ factor is interpreted as the residual wordsheet
Lorentz-invariance of the resulting string action, and the $SU(2)\times SU(2)$
factor as the maximal holonomy group of $B$. The worldsheet fields of this
effective string action are found as coefficients of the zero modes in the
harmonic expansion on $B$ of the physical fields of \eleven. We shall now
investigate what these are.

First, the five worldvolume scalars $S^{[ij]}$ produce five worldsheet
scalars. These can be identified with the spacetime coordinates
$x^\mu(\tau,\sigma)$ of the effective D=7 superstring, since wordsheet general
coordinate invariance implies that only five of these seven worldsheet fields
are physical. Next, we consider the worldvolume spinors $\lambda_+^i$, which
span a vector space of real dimension 16. Because of their
six-dimensional chirality, their $SU(2)\times SU(2)$ representations are
correlated with their worldsheet chirality: half of them are $SO(1,1)$ chiral
spinors in the $({\bf 1,2})$ representation of $SU(2)\times SU(2)$ while the
other
half are antichiral $SO(1,1)$ spinors in the $({\bf 2,1})$ representation. The
number and chirality of the worldsheet spinors therefore depends on the number
and type of covariantly constant spinors admitted by $B$, which depends on the
holonomy group of $B$. When $B=T^4$, which has a trivial holonomy, we get 8
worldsheet spinors of one chirality and 8 of the other chirality. When $B=K_3$,
which has $SU(2)$ holonomy, we get only the 8 of one chirality and none of the
other chirality. So far, the analysis closely parallels that of [\DNP] for the
$K_3$ compactification of 11-dimensional supergravity.

Finally, we turn to the two-form potential $A^+$, or rather its gauge-invariant
field strength three-form $F^+=dA^+$. The analysis required here closely
parallels that given in [\PKTc] for the $K_3$ compactification of type IIB
supergravity: in this case the three-form $F^+$ yields scalar worldsheet fields
$X^I$ via the ansatz
$$
F^+ = dX^I(\tau,\sigma) \wedge \omega_I
\eqn\twelve
$$
where $\{\omega_I\, ;\, I=1,\dots,b_2\}$ span the $b_2$-dimensional space of
closed but not exact, self-dual or anti-self-dual, two-forms on $B$; i.e. $b_2$
is the second Betti number of $B$. Since $F^+$ is self-dual in six dimensions,
the
one-forms $dX^I$ will be self-dual or anti-self-dual {\it on the string
worldsheet} according to whether $\omega_I$ is self-dual or anti-self-dual.
This
implies that the scalar $X^I$ is either left-moving or right moving, i.e.
chiral
or anti-chiral, according to whether $\omega_I$ is self-dual or anti-self-dual.
Thus, if $b_2=b_2^+ + b_2^-$, where $b_2^+$ counts the self-dual two-forms and
$b_2^-$ counts the anti-self-dual ones, then the two-form potential $A^+$ on
the
fivebrane worldsheet will produce $b_2^+$ chiral  and $b_2^-$ anti-chiral
worldsheet scalars. For $B=T^4$, we have $b_2^+=b_2^-=3$, so the two-form
potential $A^+$ of the D=11 fivebrane yields 3 worldsheet scalars of one
chirality and 3 of the other chirality or, equivalently, 3 non-chiral scalars.
These can be identified as the maps from the worldsheet to three extra
dimensions. Since the total number of physical scalars is 8, it is clear that
the
effective string theory is 10 dimensional. Since there are also 8 non-chiral
spinors we see that the $T^4$ compactified fivebrane has precisely the physical
worldsheet field content of the D=10 type IIA string.

For $B=K_3$ we have $b_2^+=3$ and $b_2^-=19$, so the two-form potential $A^+$
of
the D=11 fivebrane yields a total of 3 non-chiral worldsheet scalars, which
we can give the same interpretation as above, plus 16 additional {\it chiral}
worldsheet scalars. Thus the physical worldsheet field content of the $K_3$
compactified D=11 super-fivebrane is eight non-chiral scalars, 16 additional
chiral scalars and eight chiral spinors. This is precisely the physical
worldsheet field content of the D=10 heterotic string\foot{Strictly speaking,
we
have still to show that the spinors have opposite chirality to the 16
additional chiral scalars.}.


\chapter{The supermembrane as a compactified D=10 fivebrane}

We turn now to consider the converse of the prediction of D=7 string-membrane
duality that we have just confirmed. This is the prediction that the
$K_3$-compactified D=11 supermembrane has a solitonic interpretation as a $T^3$
compactified fivebrane soliton of the heterotic string. We earlier identified
the relevant membrane as the `neutral' one, since this corresponds to the
non-singular membrane solution of the effective D=7 action \four. The neutral
membrane is a solution of the pure D=10 supergravity theory (and is
non-singular in the string metric). As such, the physical field content of the
worldvolume action is that of a six-dimensional hypermultiplet, with 4+4
on-shell boson and fermion degrees of freedom. The worldvolume action was
constructed in [\AETW] (where its $K_3$ compactification was also briefly
discussed). The dimensional reduction of this action on
$T^3$ is that of the D=7 supermembrane, with no evidence of an 11-dimensional
origin. What we need to verify the prediction of string membrane duality is an
additional 4+4 worldsheet boson and fermion degrees of freedom, and their
worldsheet action should be that of a supersymmetric sigma model with a $K_3$
target space.

To see whether these additional degrees of freedom are present in the
relevant $T^3$ compactified heterotic fivebrane we must of course consider the
full effective D=10 action. This includes the vector supermultiplets of the
$E_8\times E_8$ or $SO(32)$ gauge group, but since they are all zero in the
neutral fivebrane solution it would appear that the inclusion of the gauge
fields
cannot help. However, we should also take into account the fact that the D=10
supergravity/YM theory is {\it not} the effective action for the heterotic
string
because it is anomalous. As Green and Schwarz showed, one must include
additional Lorentz Chern-Simons terms in the Bianchi identity for the
D=10 three-form field strength, and these must be supersymmetrized. Since the
resulting action is an infinite series in powers of the D=10 Riemann tensor,
the
complete anomaly free and supersymmetric action is unknown, but it is known
that
the neutral fivebrane requires modification. It seems likely that one must
replace the neutral fivebrane by the `symmetric' fivebrane [\CHS], since this
has the same metric and dilaton as the neutral fivebrane solution of D=10
supergravity and it is known to be an exact solution of the classical string
theory. Since the YM fields now play a role, it is possible that they or
their gaugino superpartners have zero modes, the coefficients of which would
constitute additional fields on the fivebrane's worldvolume.

In fact, the YM fields of the symmetric fivebrane are precisely those of the
`gauge' fivebrane solution [\St] for which the YM gauge field takes values in
an
$SO(3)$ subgroup of $E_8\times E_8$ or $SO(32)$. The zero modes have been
discussed in [\St,\CHSb]. Consider the $SO(32)$ case. The adjoint
representation
of $SO(32)$ has the $SO(3)\times SO(29)$ decomposition
$$
({\bf 3},{\bf 1})\oplus ({\bf 1},{\bf 406})\oplus ({\bf 3},{\bf 29})\ .
\eqn\decom
$$
The Atiyah-Singer index theorem implies a zero mode of the gaugino Dirac
opeator (in the bosonic background provided by the fivebrane solution) for each
$SO(3)$ triplet in this decomposition, of which there are a total of 30. Each
of these 30 zero modes yields, on taking into account other bosonic zero modes
required by supersymmetry, a worldvolume hypermultiplet. According to this
accounting, the physical field content of the fivebrane's worldvolume action is
30 six-dimensional hypermultiplets.

We must now dimensionally reduce on $T^3$ to find the worldvolume action of a
supermembrane. Here we must take into account that on dimensional reduction a
massless six-dimensional hypermultiplet can become a massive three-dimensional
hypermultiplet  [\STo], in which case it would not appear in the effective
worldvolume  action for the supermembrane. To see how this can happen, let
$y^i$
be the $T^3$ coordinates: a massless six-dimensional hypermultiplet coupling to
a
background $U(1)$ gauge potential $A$ yields a massless three-dimensional
hypermultiplet only if the operator $\partial/\partial y^i + iA_i$ has a zero
eigenvalue when acting on functions on $S^1$, but for generic constant values
of
$A_i$ this operator has no zero eigenvalues. To apply this observation to the
30 hypermultiplets in the ${\bf 1}\oplus{\bf 29}$ representation of $SO(29)$,
we
recall that in compactifying the heterotic string theory on $T^3$ we must do so
in such a way that the effective seven dimensional field theory is the generic
one, i.e. such that $SO(32)$ is spontaneously broken to $U(1)^{16}$. This is
done
by giving generic  expectation values to the components $A_i$ of the gauge
potentials associated with a maximal abelian subgroup of $SO(32)$. Thus, the
only
hypermultiplets of the fivebrane's worldvolume action that can be expected to
survive as massless hypermultiplets in three dimensions after wrapping the
fivebrane around the $T^3$ factor of spacetime are those which do {\it not}
couple
to the background gauge fields on $T^3$. The $29$-plet of hypermultiplets
certainly do couple to these fields, so they will not contibute to the
three-dimensional worldvolume action.

This leaves only the $SO(29)$ singlet hypermultiplet. It is unclear what
happens to it. It couples to the $SO(3)$ gauge potentials but as these are used
to construct the solution it is doubtful that they can be considered as
`background' fields. In [\St,\CHSb] this hypermultiplet was identified as the
one arising from the partial breaking of translations and supertranslations by
the
fivebrane solution. If this is right then this $SO(29)$ singlet hypermultiplet
certainly remains massless, but then the total number of physical boson and
fermion degrees of freedom would be 4+4, and we would fail to verify the
prediction of string-membrane duality. However, it seems odd that worldvolume
fields should have {\it both} a Nambu-Goldstone {\it and} a topological
interpretation. After all, the topological argument applies equally (at least
for fermion zero modes) to a non-supersymmetric theory, while the argumnt
based on symmetry breaking continues to apply in the absence of any YM
multiplets. It seems more likely that the singlet worldvolume
hypermultiplet of topological origin is an {\it addition} to the one of
Nambu-Goldstone origin. If it remains massless on $T^3$ compactification, as
seems possible, there will indeed be a total of $8+8$ boson and fermion
three-dimension worldvolume degrees of freedom, as required by string-membrane
duality. It seems a challenging problem to verify this scenario, and to
further verify that the action for the `additional' worldvolume hypermultiplet
is that of a supersymmetric three-dimensional sigma model with $K_3$ as its
target
space.


\chapter{Type IIB and D=8 Dualities}

The type IIB D=10 superstring appears to remains aloof from the tangle of
relations between the D=10 heterotic and type IIA strings and the D=11
supermembrane\foot{But see [\GR].}. Nevertheless, the analysis just carried out
for the fivebrane solution of D=11 supergravity can be repeated for the
fivebrane
solution of the D=10 N=2B supergravity. In this case the worldvolume action
[\CHSb] is based on the six-dimensional N=4 vector multiplet
$$
(S^{ii'}\; ;\; \lambda_+^i, \lambda_-^{i'}\; ;\; V)
\eqn\thirteen
$$
where $i(=1,2)$ and $i'(=1,2)$ are indices of commuting $SU(2)$ groups, i.e.
this
supermultiplet has four scalars, two SU(2)-Majorana spinors of opposite
chirality, and a one-form potential $V$. In this case, compactification on
$K_3$
yields an effective non-chiral string theory with 4 scalars and 4 spinors. This
is the field content of a six-dimensional type II superstring, with no evidence
of a 10-dimensional origin. Consequently, this is not the field content of an
anomaly free string theory. However, while the vector field $V$ does not give
any
additional worldsheet scalars, it does give a worldsheet vector. A
vector has no degrees of freedom in two dimensions but might contribute to the
conformal anomaly. Consideration of the D=10 IIB superstring therefore suggests
the existence of a six-dimensional type II string theory in which the
worldsheet conformal anomaly cancellation is achieved in this way.

Perhaps a better way to bring the type IIB superstring into the picture is by a
generalization of D=7 string-membrane duality to D=8 string-threebrane duality.
If the D=10 N=2B supergravity is compactified on $T^2$ the resulting effective
D=8 theory is that of N=2 D=8 supergravity. It is again true that there are
dual forms of the D=8 supergravity action in which one couples naturally to a
string and the other to a threebrane, and one can therefore envisage a
threebrane theory in which the string appears as a soliton. The obvious
candidate
for this string soliton is the D=10 self-dual threebrane [\HS] wrapped around a
2-torus\foot{The D=10 threebrane is geodesically complete and so would qualify
as
a soliton by the criteria that we have here been insisting on, but, in contrast
to the D=11 fivebrane, the {\it multi} D=10 threebrane is very probably
singular
[\GHT].}. The worldvolume action [\DLb] of the self-dual threebrane is based on
the four-dimensional $N=4$ vector multiplet
$$
(S^{[ij]}\, ;\, \lambda_+^i\, ;\, V)\qquad (i,j=1,2,3,4)
\eqn\fourteen
$$
where the scalars $S^{[ij]}$ are a real $SU(4)$ 6-plet, the spinor
$\lambda_+^i$
is a complex chiral $SU(4)$ 4-plet (with anti-chiral complex conjugate in the
conjugate $\bar{\bf 4}$ irrep. of $SU(4)$), and $V$ is an
$SU(4)$-singlet one-form potential. Compactification on $T^2$ yields the
physical
worldsheet field content of the type IIA superstring, which is not unexpected
in
view of the equivalence between the $T^2$ compactified type IIA and type IIB
superstrings.

There is also a solitonic threebrane of the N=2 D=8 supergravity found by
wrapping a fivebrane around the two-torus. The threebrane world volume fields
of \fourteen\ are indeed found by $T^2$ compactification of either the type IIB
or
the type IIA fivebrane worldvolume actions. It is interesting to note here
that just as the type IIA and IIB superstrings become equivalent on dimensional
reduction on $S^1$ [\ABEQUIV], so also do the worldvolume actions of their
respective fivebrane solutions!

The same $D=8$ supergravity that one finds by compactification of the type IIB
superstring on $T^2$ can also be obtained by compactification of
$D=11$ supergravity on $T^3$. If the latter is viewed as the effective theory
of a
fundamental membrane theory then the $D=8$ three-form potential also couples to
a
fundamental membrane. But the magnetic dual of a membrane in eight dimensions
is
another membrane. This dual membrane is of course the $T^3$ compactification of
the $D=11$ fivebrane. An obvious conjecture, analogous to the $D=6$
string-string
duality conjecture of [\D,\W] is a $D=8$ membrane-membrane duality. According
to
this conjecture, the dual membrane is equivalent to the original one, from
which
it follows that the $T^3$ compactification of the worldvolume action of the
$D=11$
fivebrane should yield the worldvolume action of a $T^3$ compactified $D=11$
supermebrane. This prediction of $D=8$ membrane-membrane duality is easily
checked. Taking into account that a vector potential is equivalent to a scalar
in
three dimensions, the $T^3$ compactification of the D=11 fivebrane's
worldvolume
action yields an effective three-dimensional action to which the fivebrane's
worldvolume spinors contribute 8 $SO(2,1)$ Majorana spinors and to which the
fivebrane's worldvolume two-form potential $A^+$ contributes three scalars.
Allowing for three-dimensional worldvolume diffeomorphism invariance, the five
scalars from the $S$ fields of \eleven\  can be promoted to the 8 worldvolume
fields that specify the position of a membrane's three-dimensional worldvolume
in
a D=8 spacetime. Adding to these the 3 scalars from $A^+$, we see that the full
physical field content is precisely that of the $T^3$ compactified D=11
supermembrane, as predicted.


\chapter{Comments}

The proposal that the D=10 type IIA superstring theoryis an $S^1$-compactified
D=11 supermembrane theory leads, when combined with conjectured relations
between
the heterotic and type IIA superstrings, to the conjecture that the strongly
coupled heterotic string in D=7 is equivalent to a $K_3$-compactified
supermembrane theory. This conjecture has here been shown to imply a
string-membrane duality in D=7. One implication of this duality is that from
the
membrane point of view the D=7 heterotic string is a soliton obtained by
$K_3$ compactification of the D=11 superfivebrane\foot{cf. recent similar
results
for the D=6 heterotic string [\Sb,\HStr].}. Given this,  it follows that the
worldsheet action for the heterotic string should be the effective
two-dimensional field theory obtained by $K_3$ compactification of the
six-dimensional worldvolume action of the D=11 super-fivebrane. We verified
this
prediction at the linearized level, i.e. the physical field content.
Unfortunately, it is not possible at present to check this prediction beyond
the
linearized approximation because the full D=11 fivebrane worldvolume action is
unknown; its construction is now urgently needed.

Putting together the various p-brane dualities involving the type IIA
superstring, one sees that in addition to the heterotic string emerging as a
soliton found by a $K_3$ compactification of the D=11 fivebrane, the D=10 type
IIA superstring has a similar solitonic interpretation as a
$T^4$ compactified D=11 fivebrane, as does the D=11 supermembrane as a
$T^3$ compactied D=11 fivebrane. Thus, the worldsheet actions of both the
heterotic and the type IIA superstrings, and the worldvolume action of the D=11
supermembrane appear to be effective actions arrived at by compactification of
the D=11 fivebrane's worldvolume action. From this perspective, the various
string or membrane theories are directly related to the possible choice of
compactifying space. It seems hardly possible to avoid the conclusion that the
D=11 super-fivebrane must play a fundamental role in string, and membrane,
theory;
possibly via the D=11 membrane/fivebrane duality suggested in [\HT]?

\vfill\eject
\centerline{\bf Acknowledgments}
Discussions with M.J. Duff, C.M. Hull, A. Sen and E. Witten are gratefully
acknowledged.

\refout

\end